\documentclass[preprint]{aastex}
\begin{document}
\title{A Compact Population of Red Giants in the 
Blue Compact Dwarf Galaxy UGCA~290\footnote{ 
Based on observations made with the NASA/ESA Hubble Space
Telescope, obtained at the Space Telescope Science Institute, which is
operated
by the Association of Universities for Research in Astronomy, Inc., under
NASA contract NAS 5-26555.}} 
\author{Mary M. Crone}
\affil{Skidmore College, Saratoga Springs, NY 12866, USA}
\email{mcrone@skidmore.edu}
\author{Regina E. Schulte-Ladbeck}
\affil{University of Pittsburgh, Pittsburgh, PA 15260, USA}
\email{rsl@phyast.pitt.edu}
\author{Ulrich Hopp}
\affil{Universit\"{a}tssternwarte M\"{u}nchen, M\"{u}nchen, FRG} 
\email{hopp@usm.uni-muenchen.de}
\author{Laura Greggio}
\affil{Osservatorio Astronomico di Bologna, Bologna, Italy, and
Universit\"{a}tssternwarte M\"{u}nchen, M\"{u}nchen, FRG}
\email{greggio@usm.uni-muenchen.de}

\begin{abstract}
We present HST/WFPC2 single-star photometry for the blue dwarf galaxy UGCA~290, whose
morphology is intermediate between classic iE Blue Compact Dwarfs and blue dwarfs 
which exhibit no red background sheet of older stars.  The color-magnitude diagram 
of this galaxy in V and I, extending over six magnitudes, is remarkably similar 
to that of the star-forming region in the 
iE Blue Compact Dwarf VII~Zw~403.  There is no evidence for gaps in its star-formation 
history over the last billion years, and the color of its red giant branch indicates a 
very metal-poor stellar population.  From the magnitude of the tip of the red giant 
branch, we derive a distance of 6.7 Mpc, more than twice the distance estimated
from  
the brightest blue supergiants. 
\end{abstract}


\section{Introduction}

The wide array of properties in dwarf galaxies, the most common and potentially simplest
galaxies in the universe, presents a challenge to models of structure formation.  In particular,
single-star photometry of nearby dwarfs has revealed a variety of star formation
histories, even for galaxies with similar morphology (e.g. Tolstoy 1999).  
Identifying patterns among
the various internal and external properties of dwarfs is part of 
answering fundamental  
questions about structure formation, such as whether dwarf galaxy 
formation was delayed by  
photoinization in the early universe,  
what universal events
produced the faint blue excess counts at intermediate redshifts, and 
the importance of the blowout of enriched gas 
and the infall of pristine gas from the intergalactic medium (see Kunth \& \"{O}stlin 2000
for a review.) 

One especially intriguing category of dwarf galaxy is the Blue Compact Dwarf (BCD).  This
designation actually 
comprises star forming dwarfs with a range of morphologies,
selected for blue colors and/or emission line spectra, along with compact size 
(Thuan \& Martin 1981).  
They also have low metallicities, which originally made them candidates 
for ``primeval" galaxies forming in the local universe (Searle \& Sargent 1972).  
Since then, the majority of BCDs, designated types iE and nE,  
have 
revealed relatively large, low surface-brightness 
red elliptical background sheets underlying the 
blue star-forming regions (Loose \& Thuan 1986).  HST observations have now resolved 
several of these 
background sheets into red giants (e.g. Schulte-Ladbeck, Crone, \& Hopp 1998,  
Schulte-Ladbeck et al. 2000), showing that star formation 
took place in them at least
1 Gyr ago, and possibly much earlier.  Those BCDs which do not show a red background sheet
are still candidates for primeval galaxies, a hypothesis strengthened by studies of 
abundance ratios (Izotov \& Thuan 1999, but see Kunth \& \"{O}stlin for a discussion of 
alternative explanations.)  However, recent results challenge this
hypothesis:  Pox~186, a BCD previously thought not to have a background sheet,
is now seen to have
one (Doublier et al. 2000); and even the lowest-metallicity galaxy on record,
I~Zw~18, may harbor stars at least as old as 0.5 Gyr (Aloisi, Tosi, \& Greggio 1999; 
\"{O}stlin 2000.)   
Thus, the problem of whether any BCDs are primeval is not yet solved.  Moreover,  
the typical mode of star formation in iE-type BCDs does not appear to
be the traditional picture of short, intense bursts separated by long 
quiescent periods, but a more
continuous ``gasping" mode of star formation (Schulte-Ladbeck et al. 2000.)  
It is not yet clear what mode of star formation occurs in other types of BCDs.   

The dwarf galaxy UGCA~290, which was
identified as a BCD based on its bright, compact HII emission
(Karachentsev \& Makarov 1998),  provides a key to understanding these issues. 
Until now, {\it all} the BCDs resolved as deeply as the red giant branch 
have been type iE BCDs.   
Unlike these, UGCA~290 exhibits only a slight low surface-brightness halo in 
ground-based images; its star forming region appears as two large lobes which
cover most of the visible galaxy (see Figure~1).  Therefore, the stellar
content of this galaxy can test whether the properties observed for 
deeply resolved BCDs are limited to the classic iE type or not. 

\section{Observations and reductions}

We observed UGCA~290 in August of 1999 as part of GO program
8122.  
As Figure~1 shows, we captured most of the galaxy in the WFPC2 field of view, but missed 
the edge of the low-surface brightness region to the northeast. 
We obtained data in four filters:  F814W, 
F555W, F336W, and F656N.  For F814W and F555W,
which approximate the I and V filters, we took six 1300 s exposures, 
at three dither positions, 
for a total of 7800 s in each
filter.  For F336W and F656N we obtained three 1300 s exposures at one position, 
for a total of 3900 s  in each
filter.  
In this paper we focus on the results in V 
and I.  

We combined exposures at the same pointings using CRREJ and then combined the 
dithered images using DRIZZLE onto a 1600~x~1600 pixel grid. 
Figure~1 shows the color image of the Planetary Camera (PC), showing many resolved stars.  Notice
that the galaxy is relatively transparent (see the
two red background galaxies) and free of individual HII regions. 
We masked out a few obvious background galaxies before conducting single-star photometry
with DAOPHOT.  We took the zero points from the May 1997 SYNPHOT tables and 
determined our point spread function (psf) from relatively isolated stars in our images.
The residual errors from psf fitting remain below 0.1 mag for magnitudes up
to $\sim 27.0$ in F555W and $\sim 26.3$ in F814W.  
We corrected for the small foreground extinction in this direction according to 
Schlegel et al. (1998), assuming an $R_v = 3.1$ extinction curve: $A_v = 0.046$
and $A_I=0.027$. 
We have not attempted
to correct for extinction within UGCA~290, but 
its transparancy outside major regions of star formation
suggests that it is low, at least for stars in the background sheet. 
Finally, we transformed our magnitudes into ground-based V and I following Holtzman
et al (1995). 

We estimated the completeness of our photometry by adding artificial stars to our images
using ADDSTAR and checking to see what percentages of them we could recover.  In order
to maintain the same level of crowding as in the original, we created simulated images
by adding 5\% of the observed number of resolved stars,
consistent with their observed luminosity function.  We created 200 such images for
each filter in two different regions: in the PC chip, to model the more crowded inner part
of the galaxy; and in the fourth of the WF2 chip closest to UGGA~290, to model the background sheet.
In the analysis which follows, we refer to the completeness estimates for the background
sheet. 

\section{Results}

\subsection{Stellar Content} 

Figure~2 displays the color-magnitude diagram (CMD) for UGCA~290 along with that
for the classic, well-studied iE BCD VII~Zw~403 which is only 4.4~Mpc away 
(Schulte-Ladbeck et al. 1999.)  
For UGCA~290, we   
include only stars with errors smaller than 0.2 mag in
both colors.  
Our data reduction for VII~Zw~403 is described in Schulte-Ladbeck et al. (1999);
here we show only those stars in the inner 30 arcseconds of this galaxy, which
encompasses its entire star-forming core.  
Section 3.2 explains how we set the absolute magnitude scale.  

The salient features in the CMD of UGCA~290 represent stellar populations with a range of ages.
The bright ``blue plume" at $(V-I)\sim 0$ 
contains high-mass main sequence and He-burning blue loop stars, 
the bright ``red plume" at $(V-I)\sim 1.5$ contains red supergiants 
and asymptotic giant branch stars,
the ``red tail" extending past $(V-I)>2$ contains intermediate-mass asymptotic giant
branch stars, 
and the concentration of red stars at
$(V-I)\sim 1, M_I=-3.5$  
contains low-mass red giants, along with some asymptotic giants. 
For reference, we overlay both CMDs with the $Z=0.0004$ Padova stellar evolutionary
tracks for masses 40 M$_\sun$, 15 M$_\sun$, 4 M$_\sun$, and 1 M$_\sun$ (Fagotto 
et al. 1994). 
There is also a very bright, red object with $(V-I)=4.0, M_I=-7.6$ 
which is located between the two main lobes of enhanced
star formation.  This may be a background galaxy, although it exhibits no
extended structure in our images. 

The resolved stellar contents of UGCA~290 and the core of 
VII~Zw~403 are remarkably similar. 
The star formation history 
of VII~Zw~403 was modeled by Lynds et al. (1998),
but they 
focus on the asymptotic giant branch and red giant branch 
stars in the red background sheet.  We will provide a more detailed comparison of
the young stellar populations in these two galaxies in a later paper.

\subsection{Distance and Metallicity} 

The identification of the red giant branch is especially useful in that it allows 
a reliable distance estimate;  the tip of the red giant branch (TRGB) occurs at an 
absolute magnitude $M_I\sim -4$ 
with only a small dependence on metallicity
(see Lee, Freedman, \& Madore 1993.) 
The location of our TRGB appears as a rise in the luminosity of function of 
red ($V-I > 0.8$) stars at $I=25.3 \pm 0.1$.  Guided by our work with VII~Zw~403, 
we attempted to refine this measurement by considering only stars far away from the star-forming
regions, but any benefit of a better-defined RGB was outweighed by the small
number of such stars. 
We measure the ($V-I$) color of the RGB for UGCA~290 to be $1.30 \pm 0.03$ at the TRGB 
and $1.25 \pm 0.03$ at
half a magnitude fainter than the TRGB. Following Lee et al, this indicates 
a stellar metallicity of roughly [Fe/H]$=-2.0 \pm 0.1$.  (Our stated error includes only 
uncertainty in identifying
the color from the CMD;  see Lee et al. for the limitations of this method.)  
This low value suggests the possibility of 
a very low nebular
metallicity $12+$log(O/H)$ < 7.6$ (see Figure 1 of Kunth \& \"{O}stlin 2000). 
With the approprate bolometric correction,
the absolute magnitude of the TRGB is $-3.94$.  Comparing to the observed TRGB at 
$I=25.3$, the distance modulus is $29.1 \pm 0.12$, for a distance of 6.7 Mpc. The 
error of 0.12 comes primarily from identifying the location of TRGB.   
The overall calibration of the TRGB method introduces an additional, systematic error 
of 0.18 mag. 

The previous distance estimate for UGCA~290, based on the brightest blue supergiants,
was 2.8 Mpc (Makarova et al. 1998)  Our photometry is in agreement with Makarova et al.;
our brightest three supergiants do occur at $V\sim20.6$, as they obtain from ground-based
measurements.  However, the WFPC2 reveals that these supergiants are very bright indeed, 
a full five magnitudes in $M_I$ brighter than the TRGB.   
The apparently large difference in distance estimates 
is in fact not very surprising given the uncertainties
in using the brightest supergiant method for such a faint galaxy (Greggio 1986).  
For a distance of 6.7~Mpc, the apparent magnitude found by Makarova et al. corresponds
to an abolute magnitude of $M_B=-13.4$. 

\subsection{Morphology} 

Figure 3 illustrates the difference in morphologies between VII~Zw~403 and UGCA~290 
by plotting the spatial distributions of ``young" stars and ``old" stars. 
Although most regions of the CMD can contain stars with a range of
ages, very young and old populations are fairly straightforward to identify. 
The region
with $M_I < -6$ contains stars with ages $0-100$ Myr, while the region with 
$V-I> 0.8, M_I > -4$ contains stars with ages of at least 0.5 Gyr. 
For clarity, we leave out the 
stars in other regions of the CMD. 
In order to compare the intrinsic morphologies of the two galaxies fairly, 
we include only stars brighter 
than an absolute magnitude of $M_I=-3$.  Note that in both galaxies, our 
completeness is better than 80\% for such stars. 
Our completeness is actually better for UGCA~290, despite 
its greater distance, because of our longer exposure times and dithering, and the
fact that this galaxy is simply not as crowded.  
The plots in Figure 3 are set to the same kpc scale. 
There is a significant difference between the morphologies of these two galaxies,
even keeping in mind that part of the UGCA~290 halo is missed to the northeast. 

Another difference in morphology, which is clear upon comparing the images 
directly, 
is that VII~Zw~403 exhibits
several small, bright H~II regions, while the H~II in UGCA~290 is more diffuse. 
From our continuum-subtracted F656N image, the total H$\alpha$ flux of UGCA~290 is 
$2.0\pm0.3\times 10^{-13}$ erg/s/cm$^2$, which for an extinction of 0.1~mag 
corresponds to a luminosity of $1.2\times 10^{39}$ erg/s.  This is about $60\%$ 
of the H$\alpha$ luminosity of VII~Zw~403 (Lynds et al. 1998).
It is consistent with the observed relationship between $L_{H\alpha}$ 
and $M_B$ for irregular galaxies (Hopp 1999).  For a Salpeter $0.1-100$ M$_\sun$ 
initial mass function, this luminosity 
correponds to a star formation rate of 0.0085 M$_\sun$/yr (Hunter \& Gallagher 1986). 

\section{Discussion} 

There may in fact be a large range of BCD types with very similar 
recent star formation histories.
Already, three other iE BCDs have shown similar star formation histories 
in the near-IR, with extended
blue and red plumes, asymptotic giant branches, and red giants, indicating star formation
over at least 1 Gyr with no major periods of quiescence during that time (Hopp et al. 1999).
The 
stellar content of UGCA~290 shows that these features are not limited to 
the classic iE type of BCD.  Perhaps these galaxies, 
despite the differences in their early histories
reflected in their red halos, 
have recently found themselves in a 
similar star-forming 
environment --- possibly due to infall of similar gas clouds from the intergalactic medium
combined with isolated geography.  
It is tempting to speculate on universal changes which would explain these 
similarities.   For example, perhaps  
universal conditions several billion years ago caused these galaxies to burst as 
part of the population we see
as the faint blue excess, and a certain cool-down time was needed after
that burst to allow gas to settle back into the small
gravitational potential well and again reach a critical density for star formation.
A more complete inventory of the properties of UGCA~290 and other key BCDs
is needed to answer these questions. 

Another implication of our result applies to the current debate which is raging over
the status of those BCDs which exhibit no red background sheet in ground-based CCD images. 
If the star-forming region of UGCA~290 were only slightly more extended
relative to its background sheet, 
the red giant population could easily be hidden to 
ground-based observation.  Our results support the position that 
the absence of an obvious background sheet
does {\it not} show that a BCD is undergoing its first bout of star formation. 

\acknowledgments MMC and RSL acknowledge support from HST grants 8012 and 8122. UH
acknowledges financial support from SFB 375.

\clearpage
\begin{figure} \begin{center}
\scalebox{0.8}[0.8]{\includegraphics[1.0in,3.5in][7.5in,5in]{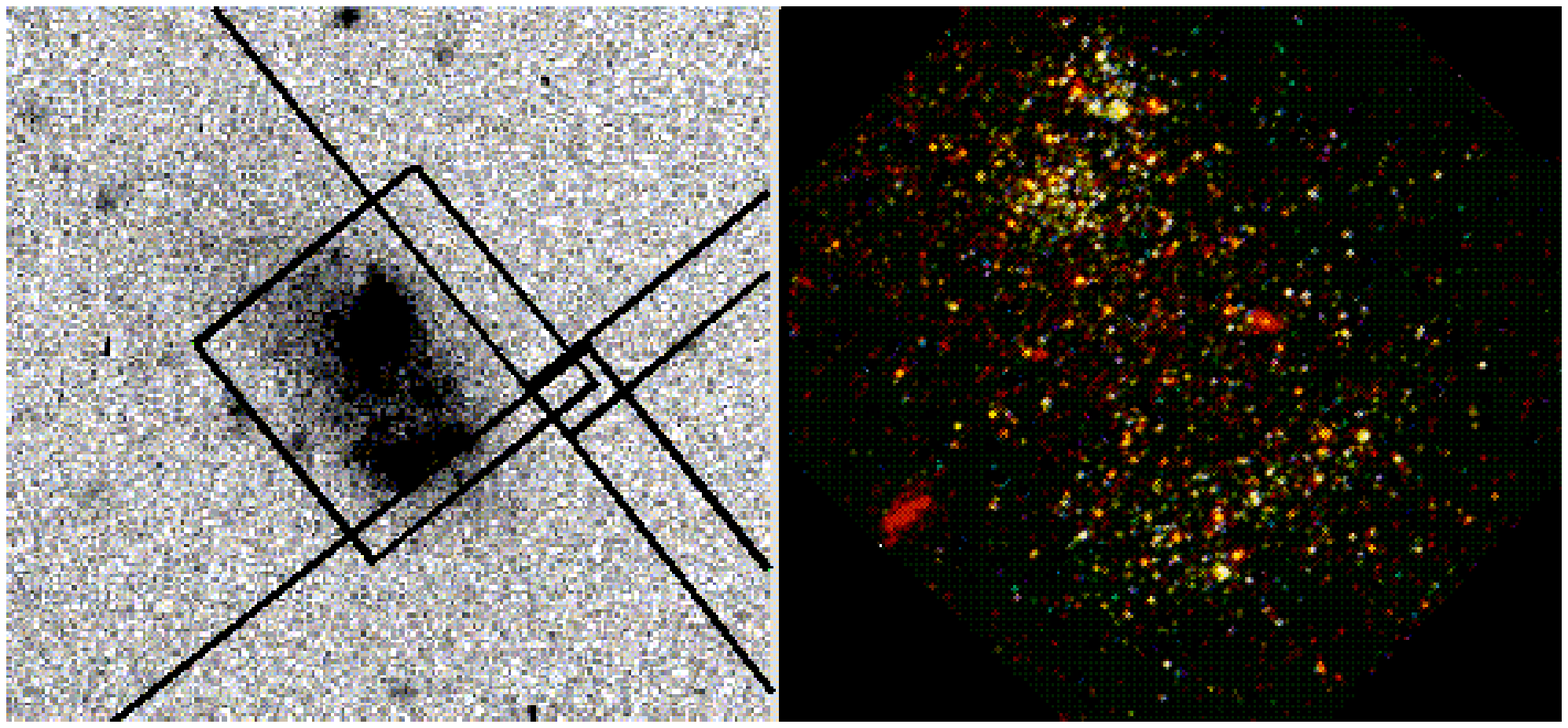}} 
\caption{WFPC2 field of view superposed on 
our 600-second R-band image taken at the Calar Alto 1.23 m
telescope (left),  
and the PC image of the galaxy center (right).  For both images, north is up
and east is to the left.} 
\end{center} \end{figure}

\clearpage
\vskip 1.0in
\begin{figure} \begin{center} 
\includegraphics[0,4.5in][8in,7in]{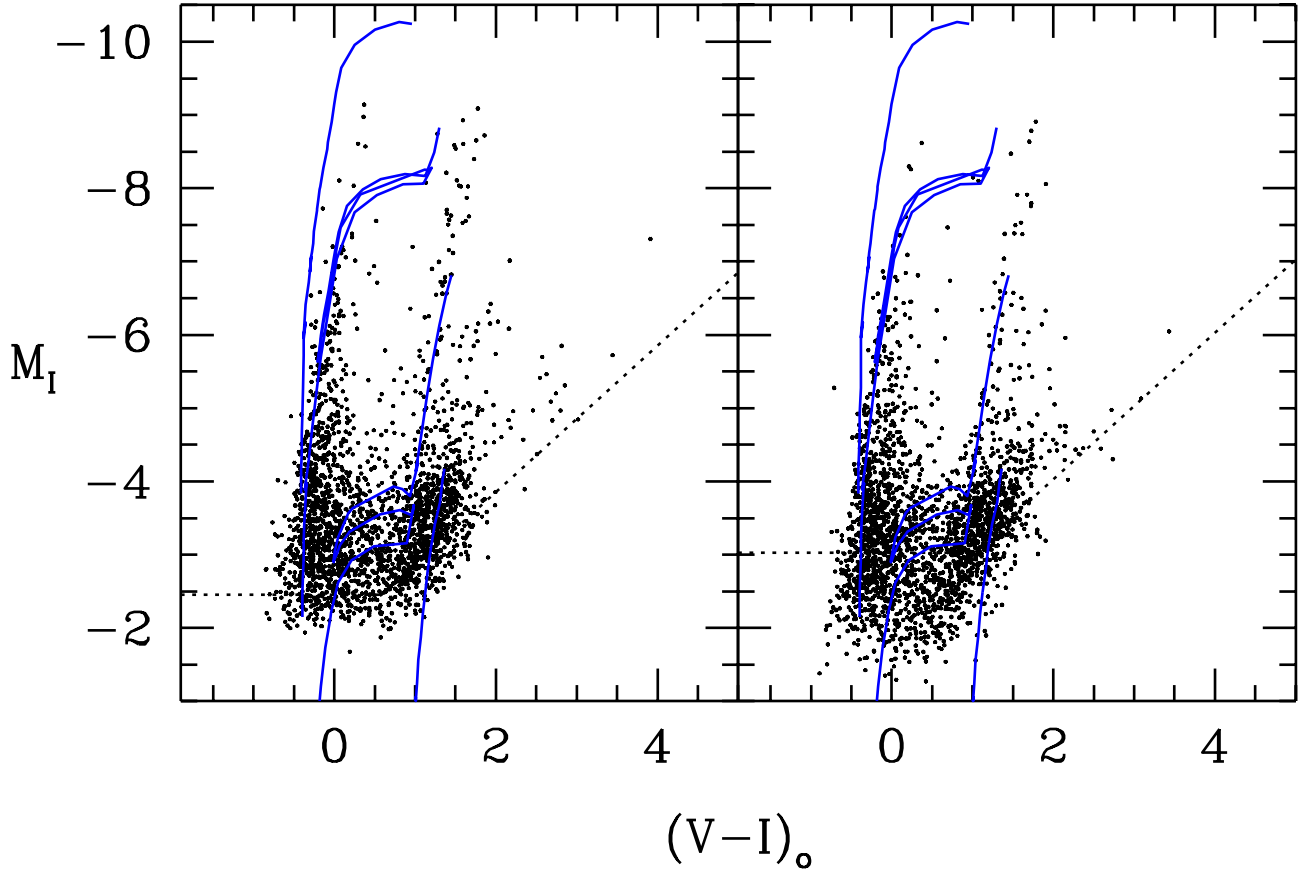}
\caption{Color-magnitude diagram of UGCA~290 and, for comparison, the inner
region of VII~Zw~403. 
Dotted lines indicate 80\% completeness limits based on artificial star tests. 
Solid blue lines indicate Z=0.0004 Padova stellar evolutionary tracks for masses
(from left to right) 40 M$_\sun$, 15 M$_\sun$, 4 M$_\sun$, and 1 M$_\sun$.  } 
\end{center} \end{figure}

\clearpage
\begin{figure} \begin{center} 
\includegraphics[0,3in][5in,5in]{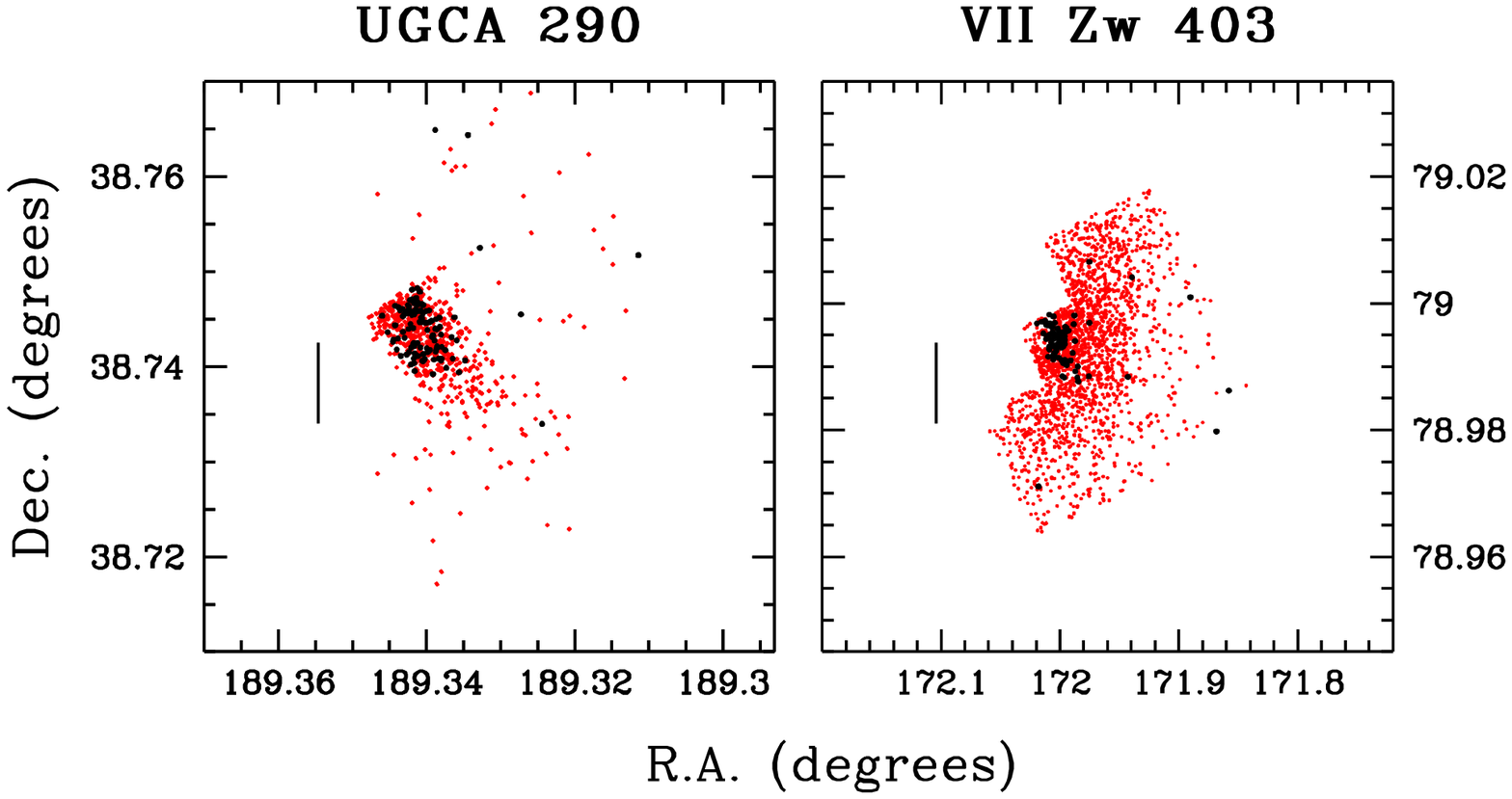}
\caption{Distribution on the sky of ``young" stars (large black dots) 
and ``old" stars (small red dots) as resolved by the WFPC2.  Both plots are set to
the same scale; the vertical lines indicate a scale of 1 kpc. 
See text for definitions of
``young" and ``old."} 
\end{center} \end{figure}

\begin{references}


\reference{ } Aloisi, A., Tosi, M, \& Greggio, L.  1999, AJ, 118, 302 

\reference{ } Arp, H. 1966, ApJS, 14, 1 

\reference{ } Doublier, V., Kunth, D., Courbin, F., \& Magain, P. 2000, A\&A, 353, 887 

\reference{ } Fagotto, F., Bressan, A., Bertelli, B. \& Chiosi, C. 1994, A\&AS, 104, 365 

\reference{ } Greggio, L. 1986 A\&A, 160, 111 

\reference{ } Holtzman, J.A., Burrows, C.J., Casertano, S., Hester, J.J., Trauger, J.T.,
Watson, A.M. \& Worthey, G. 1995, PASP, 107, 1065 

\reference{ } Hopp, U. 1999, A\&AS, 134, 317 

\reference{ } Hopp, U., Schulte-Ladbeck, R.E., Greggio, L., Crone, M.M  1999, 
in ASP Conf. Ser. 192, Spectrophotometric Dating of Stars and Galaxies, ed. I. Huberny,
S. Heap, \& R. Cornett, (ASP), 85 

\reference{ } Hunter, D.A. \& Gallagher, J.S. 1986, PASP, 98, 5 

\reference{ } Izotov, Y.I. \& Thuan, T.X. 1999, ApJ, 511, 639

\reference{ } Karachentsev, I.D. \& Makarov, D.I. 1998, A\&A 331, 891-893 

\reference{ } Kunth, D. \& \"{O}stlin, G. 2000, A\&A Rev., 10, 1 

\reference{ } Lee, M.G., Freedman, W.L., \& Madore, B.F. 1993, ApJ, 417, 553
   
\reference{ } Loose \& Thuan, T.X. 1986, in Star-Forming 
Dwarf Galaxies and Related Objects, ed. D. Kunth, T.X. Thuan, \& J.T.T. Van 
(Gif-Sur-Yvette: Editions Frontieres), 73 

\reference{ } Lynds, R., Tolstoy, E., O'Neil Jr,. E.J., \& Hunter, D.A. 1998, AJ, 116, 146 

\reference{ } Makarova, L., Karachentsev, I., Takalo, L.O., Heinamaki, P., \& Valtonen,
M., 1998, A\&AS, 128, 459-470 

\reference{ } \"{O}stlin, G. 2000, ApJ, 535, L99 

\reference{ } Schlegel, D.J., Finkbeiner, D.P., \& Davis, M. 1998, ApJ, 500, 525 

\reference{ } Schulte-Ladbeck, R.E., Crone, M.M., \& Hopp, U. 1998, ApJ, 493, L23 

\reference{ } Schulte-Ladbeck, R.E., Hopp, U., Crone, M.M., \& Greggio, L. 1999, ApJ,
525, 709 

\reference{ } Schulte-Ladbeck, R.E., Hopp, U., Greggio, L., Crone, M.M., \& Drozdovsky, I.
2000, in preparation 

\reference{ } Searle, L. \& Sargent, W.L.W. 1972, ApJ, 173, 25  

\reference{ } Thuan, T.X., \& Martin, G.E. 1981, ApJ, 247, 823 

\reference{ } Tolstoy, E. 1999, in IAU Symposium 192, eds. P. Whitelock \& 
R. Cannon (ASP), 218 

\reference{ } Vorontsov-Velyaminov, B.A. 1959, Atlas and Catalogue of Interacting
Galaxies, Part I, Moscow University 
\end{references}
\end{document}